\documentstyle[emulateapj]{article}
\def\ie{i.e.\ }
\def\eg{e.g.,\ }
\def\etal{et~al.\ }
\def\ltsima{$\; \buildrel < \over \sim \;$}
\def\simlt{\lower.5ex\hbox{\ltsima}}
\def\gtsima{$\; \buildrel > \over \sim \;$}
\def\simgt{\lower.5ex\hbox{\gtsima}}

\def\kms{km s$^{-1}$}

\def\HI{\ion{H}{1}\ }

\def\h2{H$_2$}
\def\coh2{CO/H$_2$}
\def\deg{\ifmmode^\circ\else$^\circ$\fi\ }
\def\degns{\ifmmode^\circ\else$^\circ$\fi}
\journalid{}{}
\articleid{}{}
\lefthead{Mihos}
\righthead{Gas/Star Offsets in Tidal Tails}
\slugcomment{To appear in the {\it Astrophysical Journal}}
 
\begin{document}
 
\title{The Development of Gas/Star Offsets in Tidal Tails}
 
\author{J. Christopher Mihos\altaffilmark{1}}

\altaffiltext{1}{Department of Astronomy, Case Western Reserve University,
10900 Euclid Ave, Cleveland, OH 44106, hos@burro.astr.cwru.edu}
 
\begin{abstract}

We present models of interacting galaxies in order to study the 
development of spatial offsets between the gaseous and stellar
components in tidal tails. Observationally, such offsets are
observed to exist over large scales (\eg NGC 3690; Hibbard \etal
2000), suggesting an interaction between the tidal gas and some
(unseen) hot ISM. Instead, our models show these offsets are a
natural consequence of the radially extended \HI spatial distribution 
in disk galaxies, coupled with 
internal dissipation in the gaseous component driven by the
interaction. This mechanism is most effective in systems involved in
very prograde interactions, and explains the observed gas/star offsets
in interacting galaxies without invoking interactions with a hot
ISM, starburst ionization, or dust obscuration within the tails.

\end{abstract}
 
\keywords{galaxies: interactions, galaxies: ISM, galaxies: kinematics and 
dynamics}
 
\section{Introduction}

Since the original work of Toomre \& Toomre (1972) and
Wright (1972), it has been recognized that the development of tidal
tails during galaxy encounters is predominantly a kinematic effect --
stars are tidally torn away from the parent galaxy and orbit as ``test
particles'' in the evolving potential. While self-gravity and
hydrodynamic forces can play a significant role in the development of
substructure {\it within} tidal debris (see, \eg Wallin 1990; Barnes
\& Hernquist 1992; Elmegreen, Kaufman, \& Thomasson 1993), the {\it
global} evolution of tidal tails is much simpler.  Because both gas
and stars respond to the gravitational potential in a similar fashion,
they should have similar kinematics and spatial morphology on large
scales.  To be sure, differences exist between the mass distribution
of stars and gas in tidal tails -- gaseous tails are often much more
extended than their stellar counterparts -- but this is due largely to
the differing radial distribution of stars and gas in the parent
galaxy. Because \HI disks extend further out than stellar disks (\eg
Broeils \& van Woerden 1994; Broeils \& Rhee 1997), there is more
loosely bound gas to be ejected to large distances in the tidal
debris. Nonetheless, the kinematic nature of the process predicts that
when material is tidally ejected, stars and gas should remain largely
cospatial in the resulting tidal tails.

Observationally, this situation holds true in many cases. Examples of
cospatial gas/star tails include nearly all the nearby mergers such as
the Antennae, NGC 7252, or NGC 4676 (Hibbard \& van Gorkom 1996). This
coevolutionary status of tidal gas and stars permits detailed matching
between the theoretic predictions of stellar-dynamical models and the
observed morphology and kinematics of the gas. Several investigators
have taken advantage of this fact to produce dynamical models of
specific merging systems (\eg Stanford \& Balcells 1991; Mihos \etal
1993; Hibbard \& Mihos 1995; Yun 1999; Barnes \& Hibbard 2000).

However, in a significant number of systems the \HI and
stellar tidal debris are not purely cospatial.  In many of these
cases, the observed gas/star offsets can be explained, at least
qualitatively, by interpenetrating events during galaxy
collisions. For example, the offset in the tidal bridge of NGC 7714/5
(Smith \etal 1997) is reasonably expected to occur since tidal bridges
form from material on the interface between two interacting galaxies;
collisional shocks will act to strip gas material out of the
developing tidal bridge. In retrograde encounters, short tails and
plumes form from material which passes entirely through the companion
galaxy, leading to a strong decoupling of the stars and gas (compare,
for example, Figures 11 and 12 of Mihos \& Hernquist 1996). In yet
other cases, offsets can be attributed to the fact that tidal material
has fallen back through the the merging galaxies and may be subject to
additional dissipation, as perhaps seen in the offset stellar and gaseous 
``shells'' in Centaurus A (Schiminovich \etal 1994). 

However, these explanations cannot account for a small number of cases
in which offsets exist between the gas and stars in extended tidal
tails. Notable examples include the tidal tails of NGC 520 (Hibbard,
Vacca, \& Yun 2000), 7714/5 (Smith \etal 1997), and, most
dramatically, NGC 3690 (Hibbard \& Yun 1999), in which the gas and
stars show an offset of $\sim$ 20 kpc over much of the $\sim$ 180 kpc
extent of the tail.  In these examples, none of the previously
mentioned mechanisms easily explains the observed offsets. These are
all examples of long tidal tails, which develop from the side of the
galaxy {\it away} from the companion and do not therefore come in
physical contact with the companion. Unlike the tidal shells and loops
in older mergers, the long tails here have not yet fallen back through
their central objects, and, furthermore, the length of the tails
precludes a retrograde geometry for the encounter.

At face value, these observations call into question the idea that the
evolution of tidal tails is purely kinematic, and suggest that
hydrodynamic forces may also play a role in the global evolution of
the tidal tails. Hibbard \etal (2000) suggest that the offsets may
arise from a variety of effects, including ionization of part of the
tail from a starburst wind or from the stars in the tail itself,
partial obscuration by dust in the tail, or by an interaction between
the tidal gas and a hot ISM component, either a hot gaseous halo or a
starburst wind. However, many of these solutions seem somewhat
contrived, particularly with respect to the case of NGC 3690. In this
case, the gas appears to lead the stellar tail, whereas a gas tail
expanding into a hot halo would tend to trail its stellar
counterpart. A starburst wind solution is also difficult to envision,
as the wind would need to be extremely collimated over very large
scales to result in such an extended, linear offset. While ROSAT
observations of NGC 3690 (Heckman \etal 1999) do show evidence for a
hot outflow, but the X-ray isophotes exist on scales much smaller than
needed to explain the offset tidal gas ($\sim$ 45 kpc compared to the
180 kpc extent of the tidal tail).  While the starburst wind may
extend to even larger radii at lower temperatures and densities, there
is no direct evidence for such an extended outflow.

In this paper we describe a simpler solution for the observed gas/star offsets
in some tidal tails. In this model, the offsets arise from the differing 
radial distributions of gas and stars, coupled with interaction-driven
dissipation and inflow in the tidal gas. This scenario explains the observed
offsets without resorting to esoteric processes such as hot ISM interactions,
starburst ionization, or dust obscuration. 

\section{Interaction Model}

To explore the physics which might lead to gas/star segregation in
tidal debris, we use numerical modeling of galaxy mergers to follow the
kinematic evolution of both the stellar and gaseous components of
tidal features. The models are calculated using an N-body treecode
coupled with smoothed particle hydrodynamics to follow the evolution
of the gas (TREESPH; Hernquist \& Katz 1989). The galaxy models
consist of a composite disk/bulge/halo system, in which the disk is
exponential with mass $M_d=1$ and radial scale length $h=1$, the bulge
is a flattened Hernquist model with mass $M_b=1/3$, scale radius
$a=0.4$ and axial ratio $b/a=0.5$, and the halo is a truncated
isothermal sphere of mass $M_h=5.8$, core radius $r_c=1.0$, and tidal
radius $r_t=10.0$ (see Hernquist 1993 and MH96 for more details). In
these units, one half mass rotation period for the disk is 12 units of
time, and scaling the model to the Milky Way gives length, mass, and
time scale factors of 3.5 kpc, $5.6\times10^{10} M_{\sun}$, and 13 Myr
respectively.

The model we present here is identical to the fiducial model of Mihos
\& Hernquist (1994, 1996), with one significant difference.  In those
previous models, the gas was distributed exponentially in the disk,
with a radial scale length equal to that of the stellar disk and a
total mass of $M_g=0.1M_d$.  Those models were designed to study
inflow and nuclear starbursts, and the gas distribution was chosen to
mimic the molecular gas distribution in spiral galaxies. However, the
\HI in galaxies is typically much more extended than the starlight, so
that the gas distribution used in those models was a poor
representation of the gas likely to be ejected in tidal debris. To
rectify this shortcoming, the models presented here use a different
gas mass distribution: the gas is exponential out to a radius of
$2.5h$, after which it follows a $r^{-1}$ surface density distribution
out to a truncation radius of $8h$. Because of the more slowly
declining gas distribution at large radius, Model B has a higher gas
mass: $M_g=0.25M_d$. Conceptually, Model B represents a more realistic
gas distribution; the inner exponential distribution mimics the
molecular gas content in galaxies, while the extended HI gas is
represented by the slower drop off in density at large radius. The
relatively high gas fraction in this model $M_g=0.25M_d=0.05M_{dyn}$
is more typical of late type Sb or Sc galaxies than luminous Sa's (see
compilations by Young \& Scoville 1991; Roberts \& Haynes 1994;
McGaugh \& de Blok 1997). However, given the large gas content and
high $M_g/L_B$ ratios of systems like NGC 3690 which show gas/star
offsets in their tidal debris, late-type spirals are in fact the
likely progenitors of such systems (\eg Hibbard \etal 2000).

The merger we simulate is again identical to the fiducial merger of
MH96 -- a merger between two equal mass disk galaxies, placed on an
initially parabolic orbit with pericentric distance of
$r_{peri}=2.5h$. One disk is exactly prograde ($i=0\degns$), while the
second is highly inclined to the orbital plane ($i=71\degns,
\omega=30\degns$). Figure 1 shows the evolution of the gas and stars
in the simulation. As the galaxies first pass one another, the tidal
tails are launched and the galaxies move apart, reaching a maximum
separation of $\sim 15h$ before turning around on their orbit and
merging together. In terms of the global dynamics (morphology,
kinematics, and timescales) this model evolves nearly identically to
that of MH96; more details on the global evolution and the triggering
of nuclear gas flows can be found in Mihos \& Hernquist (1994) and
MH96.

However, while the global evolution of this model is similar to that
of MH96, on smaller scales striking differences exist between the
tidal features of the two models. In MH96, the gas and stars in the
two tails are cospatial, while here they exhibit strong differences.
In the progade tail the gas is significantly offset from the stars --
and in fact leads the stellar tail -- while the inclined tail shows a
clear bifurcation in the gaseous component. Neither of these features
are seen in the MH96 tails (see Figures 3 and 4 of MH96).  The
gas-leading prograde tail in the current model captures many of the
features of the tail in Arp 299, including both the gas/star offset
and the fact that the gas and stars become cospatial again at the end
of the tail.

Figure 1 alone is enough to weaken models for gas/star offsets which
propose interactions between tail gas and a hot ISM or ICM; since no
such hot gaseous phase is present in these models, it is not a
necessary condition for gas/star segregation. Similarly, starburst
outflows are not necessary, nor is {\it in-situ} ionization of the gas
from the stellar component of the tails. In fact, given the restricted
set of physics modeled in the simulation, the offsets must be
explainable either by the simple differences in spatial distribution
of gas and stars in the progenitor disks, or by the hydrodynamic
effects which occur within the gaseous disks. We now examine the
development of the tidal tails in detail to separate these two
effects.

\section{Kinematics of Extended Mass Distributions}

To decouple hydrodynamic effects from those due to an extended disk of
material, a second model was run, identical to the first except that
all the gas particles were converted into collisionless stellar
particles.  The evolution of these collisionless ``pseudo-gas''
particles is not meant to mimic the evolution of gas, but simply to
ask whether an extended mass distribution could result in spatial
offsets from a purely kinematic evolution. We leave the question of
whether or not it could evolve that way when hydrodynamic effects are
included to the following section, which studies the full hydrodynamic
evolution of the extended gas disks.  For the current exercise, Figure 2
compares the distribution of stars and collisionless pseudo-gas in
the galaxies late in the interaction where the offset had developed
in the full hydrodynamic model (cf. the final frame of Figure 1). 

Although differences can be seen in the distribution of stars and
pseudo-gas, the two populations clearly coexist throughout the
tails. Pseudo-gas both leads and trails the stellar tail, forming a
sheath around the stellar tails. This distribution is similar to that
seen in other models of interacting systems, and in fact can be seen
in the original collision models of Toomre \& Toomre (1972), where
rings of particles from subsequentially larger initial radii formed
tails which were both longer and broader. This result was further
explored in Mihos \etal (1998), who examined the evolution of extended
test particle distributions in tidal debris, and showed in more detail
how material from large radius forms broader, more extended tidal
sheath. The difference between those models and the collisionless
model shown in Figure 2 is that in the latter model the pseudo-gas is
distributed at all radii, so that the sheath around the stellar tails
is filled in by pseudo-gas from the inner regions of the disk. The
bottom panels of Figure 2 show this more clearly, comparing the
distributions of the inner and outer pseudo-gas, where ``inner'' and
``outer'' are defined by gas initially interior or exterior to the
radius at which the initial gas distribution changed from an
exponential to a $r^{-1}$ distribution. Again, the outer pseudo-gas
does not lead the stellar tails, but rather is distributed throughout.

While the inclusion of an extended component in the galaxy
results in material which leads the stellar tails, it also produces a
trailing component -- again, the extended pseudo-gas is found in a
broader sheath surrounding the stellar component. Comparing such a
configuration to observed \HI tails is a bit difficult, as there are
no hydrodynamical effects included in this collisionless
calculation. Certainly one expects that on small scales \HI should be
more concentrated along the tails, as dissipation acts to keep the
internal velocity dispersion of the gas low. On the other hand, on
larger scales this calculation argues that a more extended gas
distribution could in principle lead to a more diffuse, extended
distribution of tidal gas than is suggested by the stellar tidal
tails (see, for example, \HI imaging of the Antennae by Hibbard
\& von Gorkom 1996).  Any deeper inference must appeal to models which include
hydrodynamic effects, which we address in the next section.  However,
what is clear is that, by itself, an extended spatial distribution of
gas does not result in the strong gas/star offset observed in the full
hydrodynamic model, or in the tidal tail of Arp 299. Hydrodynamical
effects internal to the galaxies must play a role in the formation of
these offsets.

\section{Hydrodynamical Effects}

The collisionless model demonstrates that extended mass distributions
alone cannot be the entire explanation for offset tidal debris. We now
examine the role that hydrodynamical effects play in the evolution of
the tidal debris. Naively, one might expect a simple comparison of the
models with and without gasdynamics (\ie the models of Figures 1 and
2) would reveal how gas physics have shaped the evolution of the
tail particles. In practice, however, it is not that simple -- due to
dissipation in the extended and relatively massive gas disks drives
further orbital decay, making the hydrodynamic model merge slightly
faster than the collisionless one. As a result, the tails in each
model evolve slightly differently, making a direct comparison
difficult.

To alleviate this problem and study the evolution of the tails in more
detail, we introduce the concept of ``phase space partners.'' We focus
on the prograde tail, and isolate gas and star particles within a
small piece of the tail where the gas/star offset is most dramatic
(Figure 3a). We then examine the model at a time one half rotation period
{\it before} the initial collision and locate the identified particles
in their preencounter position (Figures 3b,c). For each particle, we find the
particle from the other component nearest to it in phase space, \ie
for each gas particle $i$ we find the star particle $j$ with the
minimum phase space distance $\Delta_{ij}$ where
$$\Delta_{ij}^2 = {|\vec{x}_i - \vec{x}_j|^2 \over h^2} + 
                  {|\vec{v}_i - \vec{v}_j|^2 \over v_c^2}$$
where $h$ is the disk scale length and $v_c$ is the circular velocity. 
Because of the large particle number in the model (a total of 65,536
disk particles and 24,576 gas particles in each disk), the phase space 
separation between the chosen particles and their partners is very
small. This ensures that comparing the evolution of the gas particles and
their stellar partners are tracking nearly identical phase space
origins. By comparing the evolution of the particles and their partners
we can thus separate the stellar dynamical evolution of
the gas particles from their hydrodynamical evolution.

We begin by showing the evolution of the selected gas and star
particles which eventually form the scrutinized portion of the tail
(Figure 3d).  Before the encounter they are wound into the disk; as
the galaxies approach one another the isolated particles unwind until
the moment of perigalacticon.  At this point, gas and stars are both
spatially and kinematically coincident;
however, shortly thereafter the gas particles overtake and
lead the stellar particles, forming the offset tails. The thickness of
the tails reflects their random velocities; the compression of the gas
at perigalacticon effectively dissipates the random kinetic energy
injected by the collision, so that the gas tail is thinner than the
stellar tail.

We next compare the evolution of the gas particles with their stellar
phase space partners (Figure 3e). Again, this comparison shows how the
gaseous component would have evolved in the absence of hydrodynamic
forces.  Two things are apparent from this exercise. First, the distribution 
of gas particles is more compact than that of their stellar partners; again,
dissipation has acted to damp out random motion and cause the gas to
evolve more cohesively.  Second, the gas particles and their stellar
partners are essentially cospatial -- no offset exists between the
components. Hydrodynamic forces act to compress the gas but do not
accelerate it; there is no net force applied to the gas (\eg from
other gas particles in the simulation). The gas in this portion of
the tidal tail is
moving on an essentially collisionless trajectory.

If the tidal gas is acting collisionlessly on large scales, then why
is it offset from the collisionless stellar particles? In other words,
why is the stellar tail gas poor --
what happened to the gas that should have tracked with the stellar
tidal tails? Figure 3f shows the evolution of the stars and their
gaseous phase space partners. Shortly after collision, the gas
partners begin to fall back in towards the inner regions of their host
galaxy, unlike the stars which continue to expand outwards.  Because
of the exponential distribution of stars, the partner gas comes from
the inner regions of the disk, where it is more subject to the
dissipation which drives radial inflows (\eg MH96,
Barnes \& Hernquist 1996). The gas which would have been cospatial
with the stellar tidal tails has lost energy and angular momentum
and has decoupled from the stellar tidal debris.

Figure 4 directly compares the energy and spin angular momentum
content (with respect to the center of the parent galaxy) for the
different subsamples of tail particles. Comparing the gaseous and
stellar components, we see that, as expected due to its initially
larger radii, the gas starts out less tightly bound and with higher
angular momentum than the stars.  After the galaxies initially collide
($T\sim30 $), the tidal tails are ejected and both gas and stars end up with a
net increase in angular momentum and decrease in binding energy. The
path each component takes is markedly different, however. The stars
become transiently more bound as they fall into the potential well of
the companion, after which they gain energy through the spin-orbit
coupling of the interaction and form the tidal tail. In contrast, the
gas gains energy monotonically and moves smoothly into the tidal
tail. (Recall here that the gas and star subsets sample different
regions of initial phase space, so their evolutionary history can be
quite different even though they end up near each other in the
resulting tails.)

Comparing the energy and angular momentum for the gas and its stellar
partners, the values track very tightly together, reflecting the
largely collisionless global evolution of the tidal gas. However, we
again see significant differences between the stars and their gas
partners, in that as the the tails begin to form, the gas partners
decouple from the stellar component, losing energy and angular
momentum and flowing inwards. Note that the decoupling occurs not at
the moment of closest approach -- which might be expected if the
decoupling was due to shocks from the interpenetrating ISMs -- but
after the collision, when the galaxy develops a strong
self-gravitating response to the collisional perturbation.  This
result is consistent with the idea that gas loses angular momentum as
a result of the internal response of the disk, driven inwards by
gravitational torques from the stellar bar and spiral features (see,
\eg Barnes \& Hernquist 1991, 1996; Mihos \& Hernquist 1996). Once the
galaxies finally merge (at $T\sim 65$) further decoupling occurs as
gas is again driven inwards due to the merging process.

In summary, then, the gas/star offset observed in the models arises
due to a combination of the different spatial distribution of gas and
stars in the disk and the collisional evolution of the gaseous
component.  However, it is not the hydrodynamics of the gas present
in the tails which drives the decoupling; rather, it is the evolution
of the gas which {\it would} have been in the tails, but was driven
inwards by the collision which shapes the offset. In this sense, the
formation of the offset is again an internal response of the galaxies
involved, rather than due to subsequent evolution of the tidal
debris. Certainly the offset is not due to any hydrodynamic
interaction between the tail gas and any other gaseous component such
as a hot halo or IGM.

\section{Geometrical Effects}

The previous discussion has focussed on the prograde tail, coming
from the disk whose rotational plane matched the orbital plane of
the encounter. Since for this galaxy the encounter is exactly prograde
($i=0\degns$), the response of the disk is maximized, both in the ejection
of tidal tails and the onset of dissipational inflows. Studies of developing
caustics in tidal tails show their strong dependence on disk inclination
(Wallin 1990). In encounters where the perturbation acts entirely
in the disk plane, as in a purely prograde encounter, the developing caustics
lead to strong shocks and density enhancements in the gas. If the encounter
is more inclined, the caustics become more three dimensional, reducing
the effects of the caustic-driven shocks. It is natural therefore
to ask how sensitive our result is to varied encounter geometries, and
we assess this by examining the structure of the tail in the more
inclined ($i=71\degns$) disk. 

Figures 5a-c shows the inclined tail projected onto the disk plane of its
host galaxy. Unlike the morphology of the prograde tail, the inclined
tail shows gas distributed across the tidal tail, with a spatial
spread similar to that of the stellar component (as seen, for example,
in Figure 2).  Further inspection, however, reveals an
interesting twist to tidal tail formation -- the inclined tail
actually shows a somewhat bifurcated structure. This bifurcation is
due in part to material from the prograde companion which has been
accreted by the inclined host galaxy and has formed a secondary,
slightly trailing tidal tail. Material in this secondary tail is a
mixture of original and accreted material, while the leading tail
consists solely of material from the original disk. This mixing of
tidal debris from the two galaxies into a single tail could in
principle lead to peculiar abundance patterns in tidal tails, which
could host material with a range of abundances from different
progenitors.

Although a significant fraction of gas in the interior portion of the
inclined tail has been accreted from the prograde companion, the
majority of the gas in the tail comes from the host galaxy. In other
words, the lack of an observed gas/star offset is not because the gas
tail has been filled in via accretion, but rather because the offset
never developed in the inclined tail. An edge-on view of the inclined
tail (Figure 5d-e) shows how the tail has warped out of its original
disk plane, reducing the effects of collisional shocks at driving
inflow along the tail. The lower angular momentum gas which was easily
driven inwards in the prograde tail now remains largely in place in
the inclined tail. Geometry does thus play a role in the establishment
of offsets by altering the efficiency at which the inner gas decouples
from the developing tidal tail. This argues that tails which show a
marked offset -- such as that in Arp 299 -- probably arise from
encounters in which the host disk was very prograde.

\section{Discussion}

The fact that we can account for the offset between the gas and stars
in tidal tails without the need for a separate hot ISM component
argues against these offsets necessarily arising from interactions
with gaseous halos or starburst winds. We emphasize here that the
initial conditions are far from being contrived; on the contrary, the
radial distribution of \HI in galaxies is typically much more extended
than the stellar component (\eg Broeils \& van Woerden 1994; Broeils
\& Rhee 1997).  In truth, the extended gas disk in our model is a more
realistic initial galaxy model than those used by MH96 or Barnes \&
Hernquist 96, where purely exponential gas distributions were used. In
essence, better initial conditions have yielded a more accurate
description of the formation and evolution of the tidal tails, showing
the natural development of the gas/star offset.

These models indicate that the offset is produced when the tail
comes from a very prograde encounter; if the encounter is fairly
inclined, the dissipation in the gas component is reduced and the
offset inhibited. How does this prediction relate to the observed \HI
properties of mergers? Most nearby mergers do not show strong offsets,
consistent with the idea that the geometry must be fairly
specific to yield offsets. In two of the more extreme examples of tidal
offsets -- NGC 3690 and NGC 520 -- the orbital geometry inferred from
the measured kinematics of the system suggests a prograde encounter 
(Hibbard \etal 2000), although with significant uncertainties in the
exact geometry. If our explanation of the offset holds true,
our constraints on the geometry are even more severe than those
inferred from the global kinematics of these systems.

Our model also explains the fact that the offset gas tail in NGC 3690
appears to curve back onto the optical tail at large radius (the ``hook''
described by Hibbard \& Yun 1999). Material in the tail at large radius
comes from the most loosely bound material at initially large radius in
the progenitor disk. Coming from the outskirts of the disk,  this
material is less susceptible to the inflow driven by the self-gravitating
response of the host stellar disk. This material develops in an almost
purely collisionless evolution so that the stars and gas remain cospatial.
It is only further inwards along the tail that the offset develops, since it
is here where the tidal material would have come from deeper in the progenitor
galaxy, and be more prone to dissipative effects. 

Given the different spatial distributions of gas and stars in the
tidal tails, it is also interesting to ask if we see kinematic
segregation between the components as well. This has particularly
important ramifications for the frequent exercise of modeling specific
galaxies based on \HI or H$\alpha$ velocity fields. Because of the
increased computational cost of running hydrodynamic simulations,
collisionless N-body models are typically used to match the observed
(gas-phase) velocities (\eg Stanford \& Balcells 1991; Mihos \etal
1993; Hibbard \& Mihos 1995; Yun 1999; Barnes \& Hibbard 2000).  If
severe kinematic segregation exists in the tidal debris, models
derived from matching collisionless kinematics to gas velocities may
be in serious error.  To examine this possibility, Figure 6 shows a
simulated position-velocity plot for both the gas and stars, observed
in the plane of the prograde tail such that the portion of prograde
tail which shows the offset moves directly towards the observer.  This
viewing geometry results in the {\it maximum} observed kinematic
offset between the stars and gas in the prograde tail.  These
position-velocity plots show that the kinematic offsets are
observable, with offset amplitudes of $\sim$ 50--70\kms, but that they
do not radically affect the inferred global kinematics of the
system. While detailed model-data matching may hint at differences
between the stellar and gas kinematics (see, for example, the
discussion in Hibbard \& Mihos 1995), the derived models should not be
seriously misled by these effects.

Finally, we emphasize that our explanation of gas/star offsets in tidal
debris is not the only possible mechanism which may produce
offsets. Certainly many examples exist of kinematic decoupling which
occurs when the encounter draws tidal material through the inner
portions of galaxies, such as retrograde encounters or late infall
from tidal tails. The anti-correlation noted by Hibbard \etal (2000)
between gas and stars in the tidal debris around Arp 220 are one
possible example of this situation, where the stubby tidal features
seen in the optical are most likely not due to an extremely prograde
encounter. Also, as Hibbard \etal (2000) point out in their discussion
of gas/star offsets, many interacting galaxies possess strong
starburst winds (Heckman, Armus, \& Miley 1987, 1990) and models
suggest that collisionally heated halo gas may exist (Barnes \& Hernquist 1996); the
interaction between the tidal \HI and these hot gas components may yet
be significant in some cases. What our model shows is that the
observed offsets in long tidal tails are easily explained without the
need for a hot component, but they do not rule out the presence -- or
even the effects -- of such a hot component.

\acknowledgements

I thank Lars Hernquist and John Hibbard for many useful discussions.
This work is supported through a grant of computing time from the
San Diego Supercomputer Center, and by the National Science Foundation
through a CAREER Fellowship grant \#9876143.

%\clearpage
 
\begin{figure}
\centerline{\bf See color Figure 1.}
%\epsscale{0.8}
%\plotone{fig1rgb.ps}
\caption{The merger evolution of two equal mass disk galaxies with extended
gas disks. In this sequence, stars are shown in yellow and gas in blue (where
the gas and stars commingle at high density the colors mix to white). Note the
development of the gas/star offset in the prograde tail (to the lower right 
in the bottom panels). Time is noted in each panel; scaled to the Milky Way,
the sequence spans approximately 600 Myr.}
\end{figure}

%\clearpage
\begin{figure}
\plotone{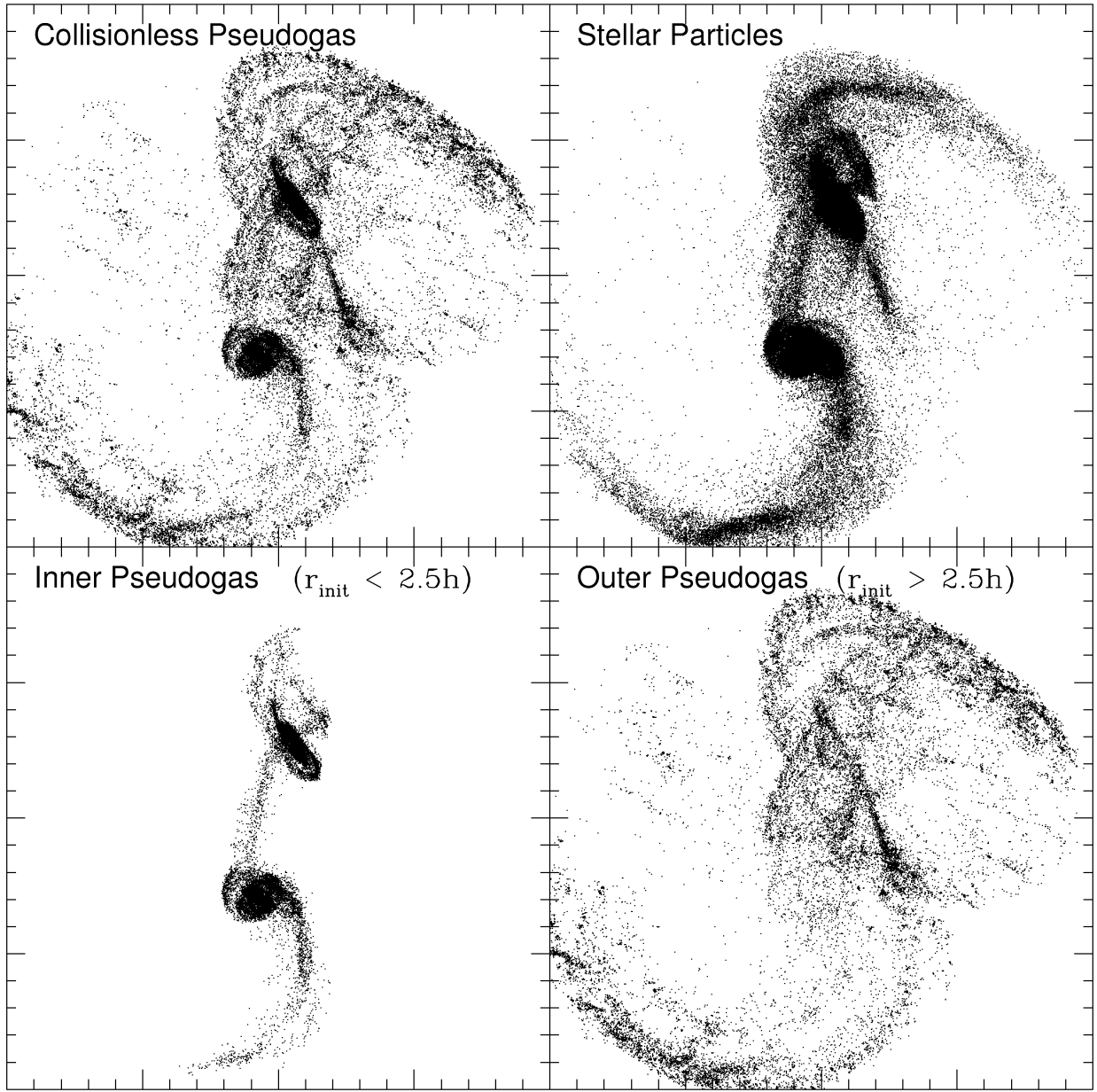}
\caption{A comparison of stellar and pseudogas morphology for a purely
collisionless model. ``Pseudogas'' refers to the gas particles in the
full hydrodynamical simulation (Figure 1) which have been converted to
collisionless particles to assess the collisionless evolution of
extended gas disks (see text). This snapshot corresponds to the final
frame in Figure 1. The bottom panels break up the pseudogas into that
coming from the inner disk and outer disk respectively, where inner
and outer is defined relative to the radius in the progenitor disks
where the gas distribution changed from exponential to $r^{-1}$. In
this exercise, no offset is seen between the stars and extended
pseudogas.}
\end{figure}

%\clearpage
\begin{figure}
\centerline{\bf See color Figure 3.}
%\epsscale{0.9}
%\plotone{fig3.ps}
\caption{The evolution of a subset of the particles in the prograde tail.
a) In a region of the prograde tail showing a strong gas/star offset, stellar
(red) and gaseous (blue) particles are identified. b) and c) These particles
are then located in the disk at a time approximately one rotation period
before the collision. d) The evolution of these particle subsamples, observed
in a frame of reference centered on the prograde disk. e) Evolution of the
gas subsample and its stellar phase space partners. f) Evolution of the
stellar subsample and its gaseous phase space partners. See text for details.}
\end{figure}

\begin{figure}
\plotone{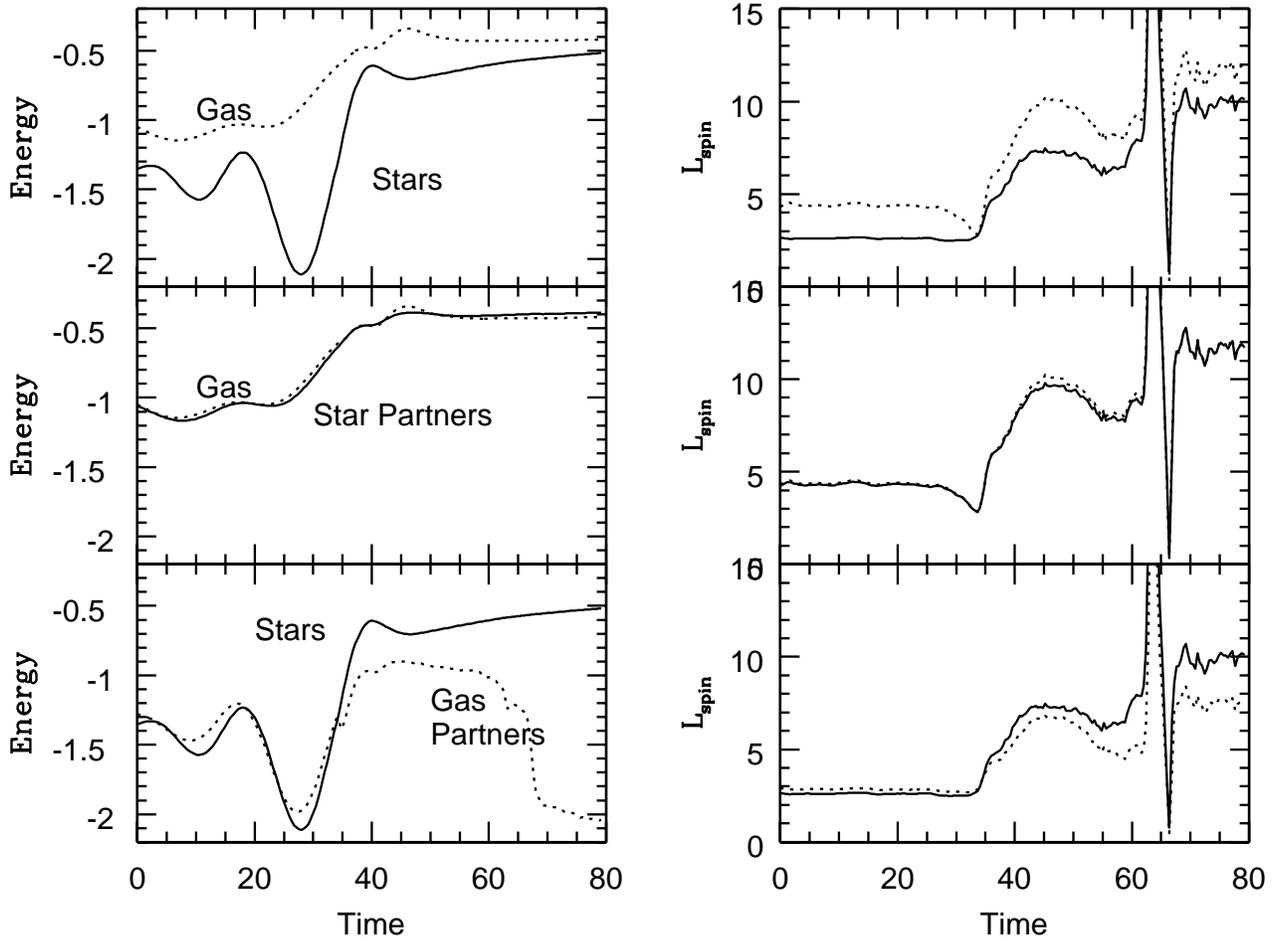}
\caption{Evolution of energy (left panels) and spin angular momenta
(right panels) of tail particle subsets. From top to bottom, the
panels compare the evolution of the gas and star subsets, the gas and
its stellar phase space partners, and the stars and their gaseous
phase space partners. In each panel, the gaseous component is shown
with a dotted line, while the stellar component is shown with a solid
line. Initial impact occurs at T $\sim$ 30, while merging happens at T
$\sim$ 65. In this figure, energy and angular momentum are measured relative
to the center of the initial disk.}
\end{figure}

\begin{figure}
\plotone{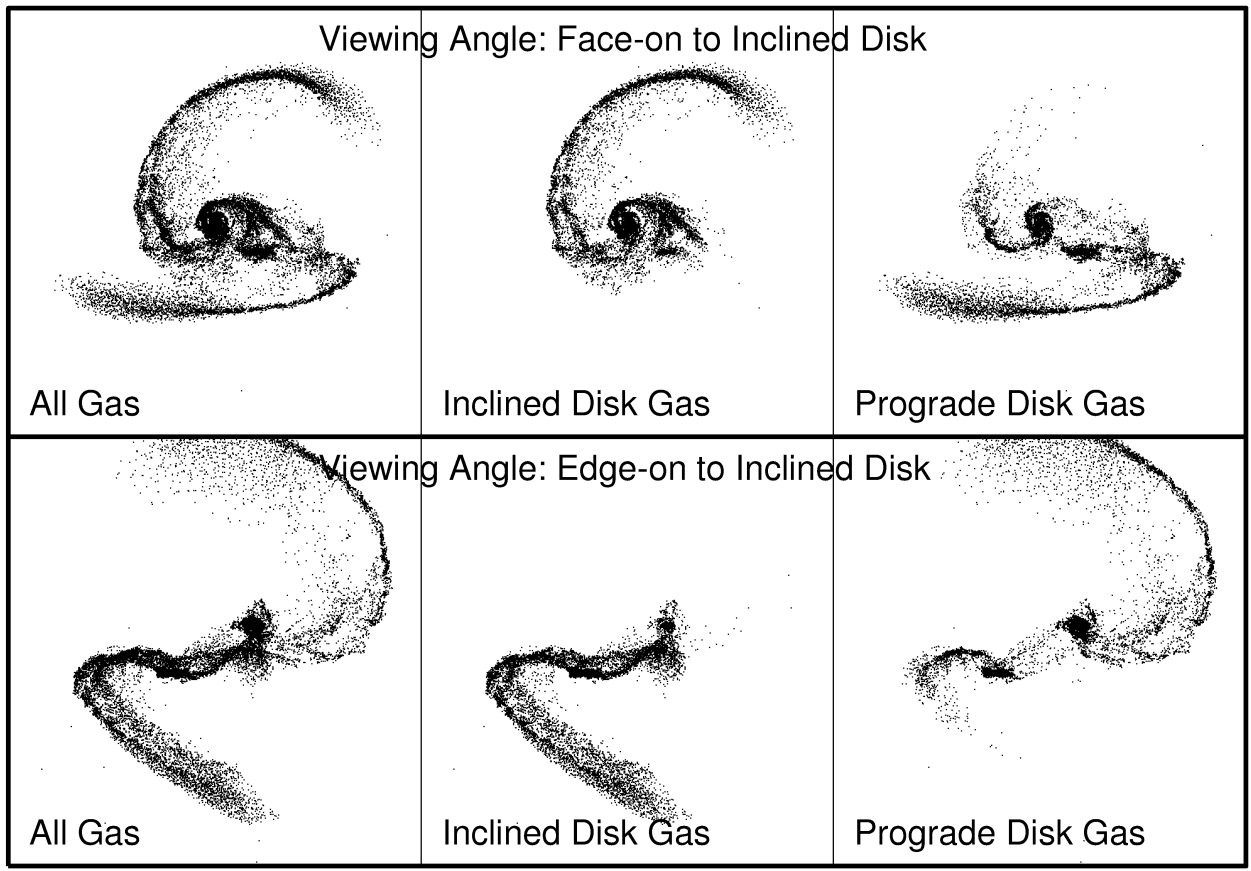}
\caption{Views of the gas in the disk initially inclined to the orbital plane.
The top panels show
the system rotated so that the inclined disk is viewed face on, while the
bottom panels show the system rotated so that the inclined disk is viewed
edge on. From left to right, the frames show all the gas, the gas initially
in the inclined disk, and the gas initially in the prograde disk. Note
the bifurcated tail, the smooth distribution of gas throughout the tail,
and the accretion of gas from the prograde disk into the inclined tail.}
\end{figure}

\begin{figure}
\epsscale{0.8}
\plotone{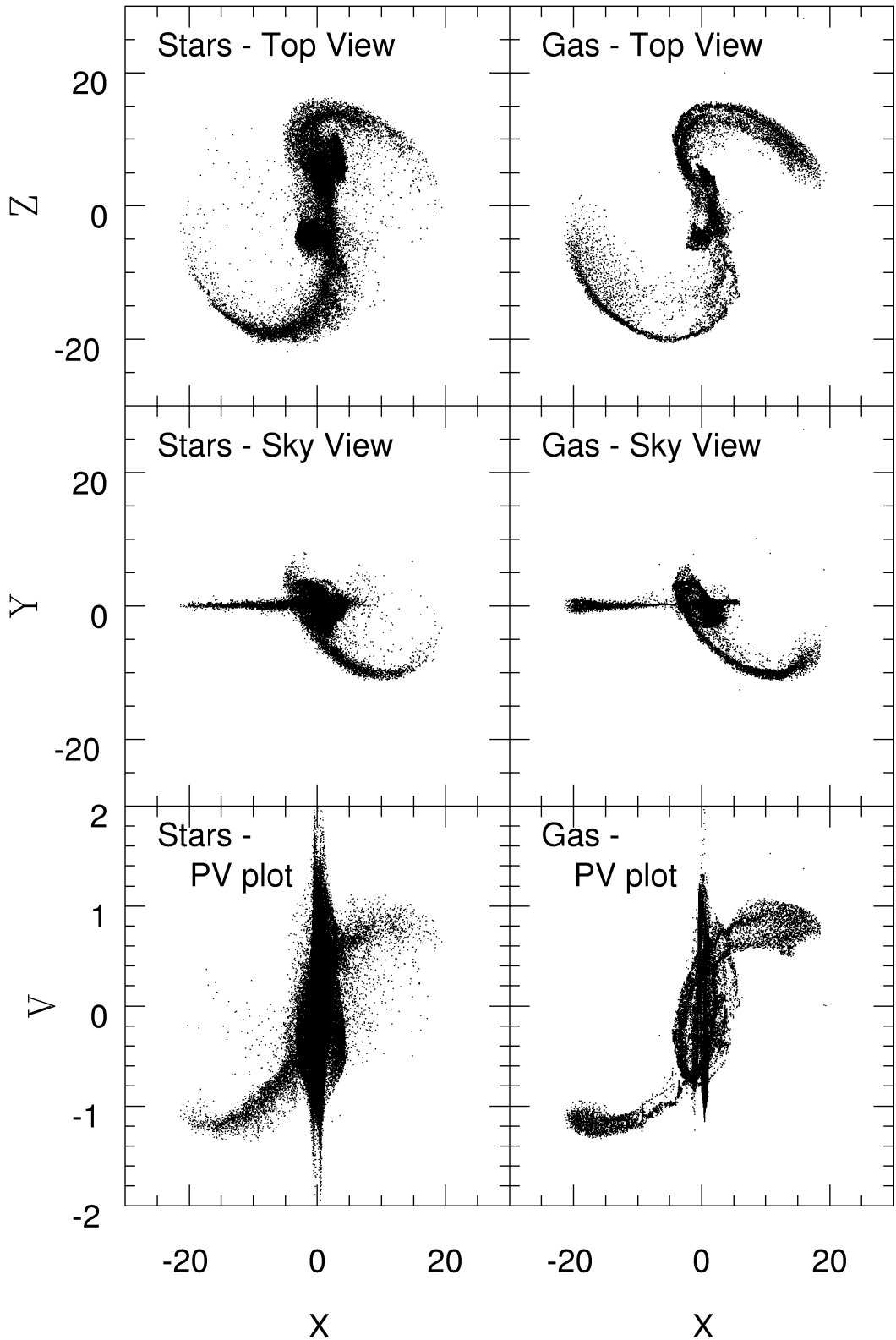}
\caption{A simulated position-velocity plot for the gas and stars in
the merger model. The model is observed late in the interaction, at
time $T=53$ when the offset is pronounced, and viewed in the plane of
the encounter, along a line of sight where the offset prograde tail is
moving towards the observer. The top panels show a top view of the
system, while the middle panels show the observed view (\ie the ``sky
plane''). The bottom panel shows the observed position-velocity
plot. Along the portion of the tail which shows the gas/star spatial
offset (X$\sim -5$), kinematic offsets are observed with magnitude
$\Delta v \sim 0.3$, or $\sim$ 60 \kms when the system is scaled to
the Milky Way.}
\end{figure}

\clearpage

\end{document}